\documentclass[notitlepage,a4paper,aps,prd,onecolumn,superscriptaddress,nofootinbib,groupedaddress,longbibliography]{revtex4-1}
\usepackage{amsmath}
\usepackage{amsfonts}
\usepackage{amssymb}
\usepackage[latin1,utf8]{inputenc}
\usepackage[T1]{fontenc}
\usepackage[colorlinks=true]{hyperref}

\newcommand{\tp}[1]{\overset{\bullet}{#1}\vphantom{#1}}
\newcommand{\lc}[1]{\overset{\circ}{#1}\vphantom{#1}}

\begin{document}

\title{Covariant formulation of scalar-torsion gravity}

\author{Manuel Hohmann}
\email{manuel.hohmann@ut.ee}
\affiliation{Laboratory of Theoretical Physics, Institute of Physics, University of Tartu, W. Ostwaldi 1, 50411 Tartu, Estonia}

\author{Laur J\"arv}
\email{laur.jarv@ut.ee}
\affiliation{Laboratory of Theoretical Physics, Institute of Physics, University of Tartu, W. Ostwaldi 1, 50411 Tartu, Estonia}

\author{Ulbossyn Ualikhanova}
\email{ulbossyn.ualikhanova@ut.ee}
\affiliation{Laboratory of Theoretical Physics, Institute of Physics, University of Tartu, W. Ostwaldi 1, 50411 Tartu, Estonia}

\begin{abstract}
We consider a generalized teleparallel theory of gravitation, where the action contains an arbitrary function of the torsion scalar and a scalar field, $f(T,\phi)$, thus encompassing the cases of $f(T)$ gravity and nonminimally coupled scalar field as subclasses.
The action is manifestly Lorentz invariant when besides the tetrad one allows for flat but nontrivial spin connection. We derive the field equations and demonstrate how the antisymmetric part of the tetrad equations is automatically satisfied when the spin connection equation holds. The spin connection equation is a vital part of the covariant formulation, since it determines the spin connection associated with a given tetrad. We discuss how the spin connection equation can be solved in general, and provide the cosmological and spherically symmetric examples. Finally we generalize the theory to an arbitrary number of scalar fields.
\end{abstract}

\maketitle

\section{Introduction}\label{sec:intro}
General relativity, which encodes the effects of gravity in the curvature, has been remarkably successful in describing a wide range of phenomena. However, to give a satisfactory account of cosmology would necessitate an \textit{ad hoc} inclusion of major extra matter ingredients responsible for triggering the rapid expansion of the early universe (inflation), for the observed structure formation and galactic rotations (dark matter), as well as for the accelerated expansion of the present universe (dark energy).
These problems keep motivating the study of extensions of general relativity, whereby many different theories and approaches have been proposed and considered \cite{Capozziello:2010zz,Clifton:2011jh,Nojiri:2017ncd}. Perhaps the most popular and promising of those are generalizing the gravitational action to be some function of the curvature scalar, $f(R)$, and the inclusion of a nonminimally coupled scalar field (scalar-tensor gravity). Such models arise naturally when quantum effects are taken into account \cite{Birrell:1982ix}, but they also happen to be favored in e.g.\ describing the spectrum of fluctuations from inflation \cite{Starobinsky:1980te,Bezrukov:2007ep,Ade:2015lrj}, and naturally permit the effective barotropic index $\mathrm{w}_{\mathrm{eff}} <-1$, as recent observations seem to suggest \cite{Ade:2015rim,Pan:2017zoh}.

From the geometric point of view curvature is not a property of spacetime \textit{per se}, but a property of the chosen connection. General relativity adopts Levi-Civita connection, which implies vanishing torsion and nonmetricity, but allows nontrivial curvature. An alternative approach, first probed by Einstein himself \cite{Sauer:2004hj}, would be to take Weitzenböck connection, which sets curvature and nonmetricity to zero, but allows nontrivial torsion. The ensuing theory where $R$ is replaced by the torsion scalar $T$ in the action is known as teleparallel equivalent of general relativity \cite{Moller:1961,Aldrovandi:2013wha,Maluf:2013gaa,Golovnev:2018red}, since its observational predictions exactly match those of general relativity. Things get more interesting, however, when one considers an extended theory, e.g.\ generalizing the action to $f(T)$ \cite{Bengochea:2008gz,Linder:2010py} or introducing and nonminimally coupled scalar field \cite{Geng:2011aj}. It turns out that the extended theories based on torsion differ from their counterparts based on curvature. This realization launched a flurry of studies regarding dark energy, inflation, black holes, and other solutions and properties of the extended teleparallel theories \cite{Cai:2015emx}.

There was a catch, though. Teleparallel gravity is usually formulated in the formalism of tetrad and spin connection, the latter being independent of the former. In the teleparallel equivalent of general relativity the spin connection does not affect the tetrad field equations, and can be chosen arbitrarily \cite{Aldrovandi:2013wha}. Interpolating this property to the extensions like $f(T)$ or scalar-torsion gravity leads to a problematic result, for the action fails to be locally Lorentz invariant~\cite{Li:2010cg,Sotiriou:2010mv}, violating the basics of the tetrad formalism. It was argued that therefore these theories implied preferred frame effects, acausality, and were inhabitated by extra spurious degrees of freedom~\cite{Li:2011rn,Ong:2013qja,Izumi:2013dca,Chen:2014qtl}. The Lorentz invariance issue is fixed in the covariant formulation of the theory~\cite{Krssak:2015oua}, which allows nontrivial spin connection compatible with vanishing curvature, i.e., flat spin connection (but the question of the degrees of freedom is still under investigation \cite{Nester:2017wau,Ferraro:2018tpu}).

After accepting nonvanishing spin connection there arises an obvious question how to determine it. An answer to the latter came only recently in the context of $f(T)$ gravity. Namely, variation of the action with respect to spin connection by carefully maintaining the flatness property yields an equation which fixes the remaining six components of the spin connection \cite{Golovnev:2017dox,Krssak:2017nlv}. This equation involves only the first derivatives of the spin connection, so one may ask whether the spin connection is an independent dynamical quantity in $f(T)$ gravity. One can not set the spin connection arbitrarily to zero, but for a given tetrad must make sure the spin connection satisfies the respective condition.
As a pleasant byproduct it turns out that when the condition on the spin connection is satisfied, the antisymmetric part of the tetrad field equations vanishes automatically \cite{Golovnev:2017dox}. It is remarkable that this feature also hold in much more general theories of torsion \cite{BeltranJimenez:2017tkd,Hohmann:2017duq}.

In the current paper we consider a generalized form of scalar-torsion gravity where the action involves an arbitrary function of the torsion scalar and a scalar field, $f(T,\phi)$, thus including both the $f(T)$ and nonminimally coupled scalar models as particular subcases. After recalling the main geometric formulae and defining the action in Sec.~\ref{sec:model}, we derive the field equations by varying the action with respect to the tetrad, flat spin connection, and the scalar field in Sec.~\ref{sec:feqs}. The spin connection equation generalizes the result found for $f(T)$ gravity \cite{Golovnev:2017dox,Krssak:2017nlv}, and shares the property that it automatically makes the antisymmetric part of the tetrad equations to be identically satisfied, as we demonstrate explicitly. In this section we also show how the spin connection equation is instrumental in guaranteeing the conservation of matter energy-momentum, and how the field configurations with constant $T$ and $\phi$ reduce the equations to those of general relativity. In Sec.~\ref{sec:connection} we discuss different possibilities how to solve the spin connection equation, and concisely present the examples of simple diagonal tetrads with associated spin connection corresponding to cosmologies of spatially flat, spherical and hyperbolic homogeneous and isotropic spacetimes, Kasner anisotropic spacetime, as well as a general static spherically symmetric spacetime. Later in Sec.~\ref{sec:multi} we develop a further generalization of the theory to multiple scalar fields, and give the respective field equations. The article ends in Sec.~\ref{sec:discussion} with a summary and discussion.

\section{Scalar-torsion model}\label{sec:model}
We start our discussion with a brief outline of the scalar-torsion model we consider in this article. In section~\ref{ssec:kinvar} we define the kinematic variables of the theory, and briefly review the definition of the terms we will use in the action. The action itself is presented in section~\ref{ssec:action}. Finally, in section~\ref{ssec:special}, we list a number of special cases of our model that have been discussed in the literature.

\subsection{Kinematic variables}\label{ssec:kinvar}
We derive our model from the covariant formulation of teleparallel gravity~\cite{Krssak:2015oua,Golovnev:2017dox}, in which the basic variables in the gravity sector are a tetrad \(h^a{}_{\mu}\) and a spin connection \(\tp{\omega}^a{}_{b\mu}\), and augment these by adding a scalar field \(\phi\). (Here the Greek indices correspond to the spacetime coordinates, while the Latin indices pertain to an orthonormal frame with Lorentzian metric $\eta_{ab}$.)
For a given spacetime metric
\begin{equation}\label{eqn:metric}
g_{\mu\nu} = \eta_{ab}h^a{}_{\mu}h^b{}_{\nu}\,,
\end{equation}
the corresponding tetrad is not defined uniquely, but only up to a local Lorentz transformation which transforms the spin connection as well,
\begin{equation}
h'^a{}_\mu =\Lambda^a{}_b  h^b{}_\mu  \,, \quad
\tp{\omega}'^a{}_{b \mu}=\Lambda^a{}_c \, \tp{\omega}^c{}_{c \mu} \, \Lambda_b{}^d + \Lambda^a{}_c \,  \partial_\mu \Lambda_b{}^c \,,
\label{eqn:Lorentz transformation}
\end{equation}
here $\Lambda_a{}^b$ is the inverse of the Lorentz transformation matrix $\Lambda^a{}_b$. The transformation \eqref{eqn:Lorentz transformation} just reflects the possibility to switch between different local observers. Demanding that the spin connection vanishes is a particular gauge choice and in general means picking a specific (class of) observer(s) among the others.
The bullet (\(\bullet\)) denotes quantities related to the teleparallel spin connection, which is chosen to be flat, i.e., having vanishing curvature,
\begin{equation}\label{eqn:curv}
\tp{R}^a{}_{b\mu\nu} = \partial_{\mu}\tp{\omega}^a{}_{b\nu} - \partial_{\nu}\tp{\omega}^a{}_{b\mu} + \tp{\omega}^a{}_{c\mu}\tp{\omega}^c{}_{b\nu} - \tp{\omega}^a{}_{c\nu}\tp{\omega}^c{}_{b\mu}\,.
\end{equation}
The spin connection defines a spacetime connection with connection coefficients
\begin{equation}
\tp{\Gamma}^{\rho}{}_{\mu\nu} = h_a{}^{\rho}\tp{D}_{\nu}h^a{}_{\mu} = h_a{}^{\rho}\left(\partial_{\nu}h^a{}_{\mu} + \tp{\omega}^a{}_{b\nu}h^b{}_{\mu}\right)\,,
\end{equation}
where \(\tp{D}_{\mu}\) is the gauge covariant Fock-Ivanenko derivative, and \(h_a{}^{\mu}\) denotes the inverse tetrad, which satisfies \(h^a{}_{\mu}h_b{}^{\mu} = \delta^a_b\) and \(h^a{}_{\mu}h_a{}^{\nu} = \delta_{\mu}^{\nu}\). The connection coefficients \(\tp{\Gamma}^{\rho}{}_{\mu\nu}\) are defined such that the total covariant derivative of the tetrad vanishes (metricity condition),
\begin{equation}
0 = \tp{\nabla}_{\mu}h^a{}_{\nu} = \partial_{\mu}h^a{}_{\nu} + \tp{\omega}^a{}_{b\mu}h^b{}_{\nu} - \tp{\Gamma}^{\rho}{}_{\nu\mu}h^a{}_{\rho}\,.
\end{equation}
Note that we adopt the convention that the last index on the connection coefficients is the ``derivative'' index, while the first pair of indices are the ``endomorphism'' indices. This connection in general has non-vanishing torsion
\begin{equation}\label{eqn:torsion}
T^{\rho}{}_{\mu\nu} = \tp{\Gamma}^{\rho}{}_{\nu\mu} - \tp{\Gamma}^{\rho}{}_{\mu\nu}\,.
\end{equation}
We further use an open circle (\(\circ\)) to denote quantities related to the Levi-Civita connection \(\lc{\nabla}_{\mu}\) of the metric~\eqref{eqn:metric}, whose connection coefficients are given by
\begin{equation}
\lc{\Gamma}^{\rho}{}_{\mu\nu} = \frac{1}{2}g^{\rho\sigma}\left(\partial_{\mu}g_{\sigma\nu} + \partial_{\nu}g_{\mu\sigma} - \partial_{\sigma}g_{\mu\nu}\right)\,.
\end{equation}
In contrast to the teleparallel connection, it has vanishing torsion, but in general non-vanishing curvature. The difference
\begin{equation}\label{eqn:contor}
K^{\rho}{}_{\mu\nu} = \tp{\Gamma}^{\rho}{}_{\mu\nu} - \lc{\Gamma}^{\rho}{}_{\mu\nu} = \frac{1}{2}\left(T_{\mu}{}^{\rho}{}_{\nu} + T_{\nu}{}^{\rho}{}_{\mu} - T^{\rho}{}_{\mu\nu}\right)
\end{equation}
is called the contortion tensor. For convenience, we will denote the partial derivatives using the comma notation, e.g., \(\partial_{\mu}\phi \equiv \phi_{,\mu}\).

\subsection{Action}\label{ssec:action}
We now come to the action of our model, which we write in the form
\begin{equation}\label{eqn:action}
S = S_g[h^a{}_{\mu}, \tp{\omega}^a{}_{b\mu}, \phi] + S_m[h^a{}_{\mu}, \chi]\,,
\end{equation}
where \(S_g\) denotes the gravitational part, while \(S_m\) denotes the matter part, and matter fields are collectively denoted by \(\chi\). For the gravitational part we choose the action
\begin{equation}\label{eqn:gravaction}
S_g = \frac{1}{2\kappa^2}\int_M d^4x \, h\left[f(T,\phi) + Z(\phi)g^{\mu\nu}\phi_{,\mu}\phi_{,\nu}\right]\,,
\end{equation}
which depends on two free functions \(f\) and \(Z\), while $2\kappa^2=16 \pi G_N$ sets the Newtonian gravitational constant. Here \(h = \det(h^a{}_{\mu})\) denotes the determinant of the tetrad, while \(T\) is the torsion scalar defined as
\begin{equation}\label{eqn:torsscal}
T = \frac{1}{2}T^{\rho}{}_{\mu\nu}S_{\rho}{}^{\mu\nu}\,,
\end{equation}
with the superpotential
\begin{equation}\label{eqn:suppot}
S_{\rho}{}^{\mu\nu} = K^{\mu\nu}{}_{\rho} - \delta_{\rho}^{\mu}T_{\sigma}{}^{\sigma\nu} + \delta_{\rho}^{\nu}T_{\sigma}{}^{\sigma\mu}\,.
\end{equation}
Here the reader should be alerted that we use a convention where the definition of the torsion scalar~\eqref{eqn:torsscal} contains a factor \(\tfrac{1}{2}\), while the superpotential~\eqref{eqn:suppot} does not carry such a factor~\cite{Golovnev:2017dox}. Other authors, often in the field of $f(T)$ gravity and cosmology, use a different convention by including the factor \(\tfrac{1}{2}\) into the definition of the superpotential and leave the torsion scalar without (e.g.,~\cite{Linder:2010py}). Finally, in the literature on the teleparallel equivalent of general relativity there is yet another convention which puts the factor \(\tfrac{1}{2}\) directly into the gravitational action, while keeping the definitions of the torsion scalar and superpotential free of it~\cite{Aldrovandi:2013wha,Krssak:2015oua}. One should be careful, as these choices affect the respective factors in the field equations as well.

The gravitational part \(S_g\) of the action is complemented by a matter part \(S_m\), which is assumed to depend only on the tetrad \(h^a{}_{\mu}\) and a set of matter fields \(\chi\), whose precise content is not relevant for the purpose of this article. The only relevant quantity is the energy-momentum tensor \(\Theta_a{}^{\mu}\) defined from the variation of the matter action with respect to the tetrad,
\begin{equation}\label{eqn:enmomtens}
\delta_hS_m = -\int_M d^4x \, h\Theta_a{}^{\mu}\delta h^a{}_{\mu}\,.
\end{equation}
We demand that \(S_m\) is invariant under local Lorentz transformations \eqref{eqn:Lorentz transformation}, which then implies that the energy-momentum tensor is symmetric~\cite{Aldrovandi:2013wha},
\begin{equation}
0 = \Theta_{[\mu\nu]} = h^a{}_{[\mu}g_{\nu]\rho}\Theta_a{}^{\rho}\,.
\end{equation}
Further, we demand that \(S_m\) is invariant under diffeomorphisms, which implies that \(\Theta_{\mu\nu}\) is covariantly conserved with respect to the Levi-Civita connection,
\begin{equation}
\lc{\nabla}_{\mu}\Theta^{\mu\nu} = 0\,.
\end{equation}
We will discuss this aspect further in section~\ref{ssec:enmomcons}.

\subsection{Special cases}\label{ssec:special}
Let us note that the action \eqref{eqn:gravaction} encompasses several previously studied scalar-torsion theories as subclasses, e.g.,\
\begin{itemize}
\item teleparallel equivalent of general relativity with a minimally coupled scalar field (quintessence),
\begin{equation}
\label{eqn:teleparallel quintessence}
f(T,\phi)=T - 2 \kappa^2V(\phi) \,, \qquad
Z(\phi)=\kappa^2\,;
\end{equation}
\item $f(T)$ gravity \cite{Bengochea:2008gz,Linder:2010py},
\begin{equation}
\label{eqn:f(T) gravity}
f(T,\phi)=f(T)  \,, \qquad
Z(\phi)=0\,;
\end{equation}
\item minimally coupled scalar field in $f(T)$ gravity \cite{Chakrabarti:2017moe}
\begin{equation}
\label{eqn:teleparallel f(T) quintessence}
f(T,\phi)=f(T) - 2 \kappa^2V(\phi) \,, \qquad
Z(\phi)=\kappa^2\,;
\end{equation}
\item teleparallel dark energy \cite{Geng:2011aj}
\begin{equation}
\label{eqn:teleparallel dark energy}
f(T,\phi)=(1+2\kappa^2\, \xi \phi^2) T - 2 \kappa^2V(\phi) \,, \qquad
Z(\phi)=\kappa^2\,;
\end{equation}
\item generalized teleparallel dark energy \cite{Otalora:2013tba}
\begin{equation}
\label{eqn:generalized teleparallel dark energy}
f(T,\phi)=\left(1+2\kappa^2\, \xi f(\phi) \right) T - 2 \kappa^2V(\phi) \,, \qquad
Z(\phi)=\kappa^2\,;
\end{equation}
or \cite{Jamil:2012vb}
\begin{equation}
\label{eqn:generalized teleparallel dark energy Myrzakulov}
f(T,\phi)=\left(1+2\kappa^2\, \xi f(\phi) \right) F(T) - 2 \kappa^2V(\phi) \,, \qquad
Z(\phi)=\kappa^2\,;
\end{equation}
\item Brans-Dicke-like action with constant kinetic term coupling \cite{Izumi:2013dca}
\begin{equation}
\label{eqn:Brans-Dicke-like omega constant}
f(T,\phi)=f(\phi) T - 2 \kappa^2V(\phi) \,, \qquad
Z(\phi)=\omega\,;
\end{equation}
or with dynamical kinetic term coupling \cite{Chen:2014qsa}
\begin{equation}
\label{eqn:Brans-Dicke-like omega}
f(T,\phi)=\phi T - 2 \kappa^2V(\phi) \,, \qquad
Z(\phi)=\frac{\omega(\phi)}{\phi} \,;
\end{equation}
\item and finally the scalar-torsion equivalent to $F(T)$ gravity \cite{Izumi:2013dca}
\begin{equation}
\label{eqn:F(T) gravity equivalent}
f(T,\phi)=\frac{dF}{d\phi} T - \phi \frac{dF}{d\phi} - F(\phi) \,, \qquad
Z(\phi)=0\,.
\end{equation}
\end{itemize}
The actions \eqref{eqn:generalized teleparallel dark energy}-\eqref{eqn:Brans-Dicke-like omega} can be transformed into each other by a suitable redefinition of the scalar field $\phi$ \cite{Chen:2014qsa,Jarv:2015odu}. In the last case~\eqref{eqn:F(T) gravity equivalent} the scalar field is actually a nondynamical auxiliary field~\cite{Izumi:2013dca}.
%Of course, also the action of \(F(T)\) gravity itself is already trivially of the form~\eqref{eqn:action} with \(f(T,\phi) = F(T)\) and \(Z(\phi) = 0\). Also one may minimally couple a scalar field to \(F(T)\) gravity by setting \(f(T,\phi) = F(T) + V(\phi)\) and leaving \(Z(\phi)\) arbitrary; this is the most general case such that \(f_{T\phi} = 0\).
For the theories based only on the torsion scalar \eqref{eqn:torsscal} and without introducing derivative couplings between the scalar field and torsion \cite{Kofinas:2015hla} or a nonstandard kinetic term for the scalar field \cite{Banijamali:2012nx}, the action \eqref{eqn:gravaction} is in the most general form. As a remark, we note that in contrast to the scalar-curvature theories, the scalar torsion action \eqref{eqn:gravaction} is not invariant under conformal rescalings of the tetrad, for these introduce a coupling between the scalar field and vector torsion \cite{Bamba:2013jqa}, leading to a much broader class of theories \cite{Bahamonde:2015hza, Bahamonde:2017wwk}.

\section{Equations of the theory}\label{sec:feqs}

With the preliminaries in place we will now go on to vary the action \eqref{eqn:gravaction} with respect to tetrad in section~\ref{ssec:feqtet}, flat spin connection in section~\ref{ssec:feqcon}, and the scalar field in section~\ref{ssec:feqscal}, and write down the ensuing field equations. In section~\ref{ssec:feqrel}, we show in detail that the condition arising from the variation with respect to the spin connection is equivalent to the antisymmetric part of the tetrad field equations. In section~\ref{ssec:enmomcons}, we demonstrate that combining the equations also leads to an expression for the matter energy-momentum conservation, and the spin connection equation is instrumental in guaranteeing that. Finally, in section~\ref{ssec:grred} we explicate how the symmetric part of the tetrad field equations reduces to general relativity for configurations with a constant torsion scalar and scalar field.

\subsection{Tetrad field equation}\label{ssec:feqtet}
We start with the derivation of the tetrad field equation, which is obtained by variation of the action~\eqref{eqn:action} with respect to the tetrad \(h^a{}_{\mu}\). From the variation of the gravitational part~\eqref{eqn:gravaction} we obtain
\begin{equation}
\begin{split}
\delta_hS_g &= \frac{1}{2\kappa^2}\int_Mh\left\{\left[(f + Zg^{\rho\sigma}\phi_{,\rho}\phi_{,\sigma})h_a{}^{\mu} - 2S^{\rho\sigma\mu}T_{\rho\sigma\nu}h_a{}^{\nu}f_T - 2Zg^{\mu\nu}\phi_{,\nu}\phi_{,\rho}h_a{}^{\rho}\right]\delta h^a{}_{\mu} - 2f_TS_{\rho}{}^{\mu\nu}h_a{}^{\rho}\tp{D}_{\nu}\delta h^a{}_{\mu}\right\}d^4x\\
&= \frac{1}{2\kappa^2}\int_Mh\left[(f + Zg^{\rho\sigma}\phi_{,\rho}\phi_{,\sigma})h_a{}^{\mu} - 2S^{\rho\sigma\mu}T_{\rho\sigma\nu}h_a{}^{\nu}f_T - 2Zg^{\mu\nu}\phi_{,\nu}\phi_{,\rho}h_a{}^{\rho} + \frac{2}{h}\tp{D}_{\nu}\left(hf_TS_{\rho}{}^{\mu\nu}h_a{}^{\rho}\right)\right]\delta h^a{}_{\mu}d^4x\\
&= \frac{1}{2\kappa^2}\int_Mh\left[(f + Zg^{\rho\sigma}\phi_{,\rho}\phi_{,\sigma})g_{\mu\nu} + 2S^{\rho\sigma}{}_{\mu}\left(K_{\rho\nu\sigma} - T_{\rho\sigma\nu}\right)f_T - 2Z\phi_{,\mu}\phi_{,\nu} + 2\lc{\nabla}_{\rho}\left(f_TS_{\nu\mu}{}^{\rho}\right)\right]g^{\mu\tau}h_a{}^{\nu}\delta h^a{}_{\tau}d^4x\,.
\end{split}
\end{equation}
From the second line one can find the left hand side of the tetrad field equations with one Lorentz index and one spacetime index; however, we skip this step here and proceed with the third line, which is already written in lower spacetime indices only. Using the definitions~\eqref{eqn:suppot} of the superpotential and~\eqref{eqn:contor} of the contortion, as well as the energy-momentum tensor~\eqref{eqn:enmomtens}, the resulting field equation can be written as
\begin{multline}\label{eqn:tetradfeq}
\frac{1}{2}fg_{\mu\nu} + \lc{\nabla}_{\rho}\left(f_TS_{\nu\mu}{}^{\rho}\right) + \frac{1}{2}f_T\left(T^{\rho}{}_{\rho\sigma}T^{\sigma}{}_{\mu\nu} + 2T^{\rho}{}_{\rho\sigma}T_{(\mu\nu)}{}^{\sigma} - \frac{1}{2}T_{\mu\rho\sigma}T_{\nu}{}^{\rho\sigma} + T_{\mu\rho\sigma}T^{\rho\sigma}{}_{\nu}\right)\\
- Z\phi_{,\mu}\phi_{,\nu} + \frac{1}{2}Zg_{\mu\nu}g^{\rho\sigma}\phi_{,\rho}\phi_{,\sigma} = \kappa^2\Theta_{\mu\nu}\,.
\end{multline}
Let us remark that contrary to the $f(R)$ theories these equations contain derivatives of the tetrad (or metric) only up to the second order. Therefore several issues and subtleties characteristic of $f(R)$ gravity~\cite{Capozziello:2010zz} do not arise. This nice feature has been one of the motivations to study $f(T)$ gravity \cite{Cai:2015emx} and persists in the more general $f(T,\phi)$ models considered here as well.

In analogy to the teleparallel equivalent of general relativity \cite{Aldrovandi:2013wha}, one may notice that the left hand side of the tetrad field equations allows for the definition of a conserved gravitational energy-momentum pseudotensor. Given a minimally coupled scalar field, this can further be split into contributions from the tetrad and the scalar field. However, in the case of nonminimal coupling the energy-momenta of the tetrad and scalar fields is entwined, since the function $f$ contains both of them.

\subsection{Connection field equation}\label{ssec:feqcon}
We now come to the variation of the action~\eqref{eqn:action} with respect to the spin connection \(\tp{\omega}^a{}_{b\mu}\). Here we follow the constrained variation prescription~\cite{Golovnev:2017dox}, where the variation is restricted to a gauge covariant derivative of the form \(\delta\tp{\omega}^{ab}{}_{\mu} = -\tp{D}_{\mu}\lambda^{ab}\) with \(\lambda^{(ab)} = 0\), such that the varied spin connection remains flat, \(\delta_{\omega}\tp{R}^a{}_{b\mu\nu} = 0\). Note that the spin connection enters the action~\eqref{eqn:gravaction} only through the torsion scalar, whose variation can be written in the very simple form~\cite{Golovnev:2017dox}
\begin{equation}
\delta_{\omega}T = \delta_{\omega}\left(-\lc{R} + 2\lc{\nabla}_{\mu}T_{\nu}{}^{\nu\mu}\right) = 2\lc{\nabla}_{\mu}\delta_{\omega}T_{\nu}{}^{\nu\mu} = 2\lc{\nabla}_{\mu}\left(h_a{}^{\nu}h_b{}^{\mu}\delta\omega^{ab}{}_{\nu}\right)\,,
\end{equation}
using the relation between the torsion scalar \(T\) and the Ricci scalar \(\lc{R}\), and the fact that the latter depends only on the tetrad, but not on the teleparallel spin connection. We then obtain the variation of the gravitational part of the action as
\begin{equation}
\begin{split}
\delta_{\omega}S_g &= \frac{1}{2\kappa^2}\int_Mhf_T\delta_{\omega}Td^4x\\
&= -\frac{1}{\kappa^2}\int_Mhf_T\lc{\nabla}_{\mu}\left(h_a{}^{\nu}h_b{}^{\mu}\tp{D}_{\nu}\lambda^{ab}\right)d^4x\\
&= \frac{1}{\kappa^2}\int_Mh\partial_{\nu}f_Th_a{}^{\mu}h_b{}^{\nu}\tp{D}_{\mu}\lambda^{ab}d^4x\\
&= \frac{1}{\kappa^2}\int_Mh\left(-\tp{\nabla}_{\mu}\tp{\nabla}_{\nu}f_T + K^{\rho}{}_{\mu\rho}\partial_{\nu}f_T\right)h_a{}^{\mu}h_b{}^{\nu}\lambda^{ab}d^4x\,,
\end{split}
\end{equation}
where we have twice performed integration by parts. Due to the antisymmetry of \(\lambda^{ab}\), as well as the relation
\begin{equation}
\tp{\nabla}_{\mu}\tp{\nabla}_{\nu}f_T - \tp{\nabla}_{\nu}\tp{\nabla}_{\mu}f_T = T^{\rho}{}_{\nu\mu}\partial_{\rho}f_T
\end{equation}
for the commutator of covariant derivatives acting on the scalar function \(f_T\), the field equation reads
\begin{equation}\label{eqn:connfeq}
0 = -\tp{\nabla}_{[\mu}\tp{\nabla}_{\nu]}f_T + \partial_{[\nu}f_TK^{\rho}{}_{\mu]\rho} = \frac{3}{2}\partial_{[\rho}f_TT^{\rho}{}_{\mu\nu]}\,.
\end{equation}
In the last term the notation means that one first needs to antisymmetrize with respect all three lower indices and then sum over with the repeating upper index.
By expanding the torsion into the tetrad and the spin connection, one can also write this equation as
\begin{equation}
\label{eqn:connection condition}
\partial_{\mu}f_T\left[\partial_{\nu}\left(hh_{[a}{}^{\mu}h_{b]}{}^{\nu}\right) + 2hh_c{}^{[\mu}h_{[a}{}^{\nu]}\tp{\omega}^c{}_{b]\nu}\right] = 0\,,
\end{equation}
which is of the same form as the corresponding equation in \(f(T)\) gravity~\cite{Golovnev:2017dox,Krssak:2017nlv}. However, note that in the scalar-torsion model the function \(f\) also depends on the scalar field \(\phi\), so that the derivative reads
\begin{equation}
\label{eqn:connection condition f_T written out}
\partial_{\mu}f_T = f_{TT}\partial_{\mu}T + f_{T\phi}\partial_{\mu}\phi\,,
\end{equation}
whereas the second term is not present in \(f(T)\) gravity. This second term vanishes if and only if the scalar field is minimally coupled, \(f_{T\phi} = 0\), or in a field configuration with uniform scalar field.

The equation \eqref{eqn:connection condition} contains only the first derivatives of the spin connection which appear since we are taking the derivatives of the torsion scalar \eqref{eqn:torsscal} in $f_T$. Therefore this equation can be interpreted as a condition to determine the spin connection components associated with a given tetrad. It is a feature characteristic of the generalized teleparallel framework, since in the $f(R)$ and scalar-curvature theories the spin connection is completely determined by the tetrad via the Levi-Civita prescription.

\subsection{Scalar field equation}\label{ssec:feqscal}
Finally, we come to the scalar field equation. Variation of the gravitational part~\eqref{eqn:gravaction} of the action with respect to the scalar field yields
\begin{equation}
\begin{split}
\delta_{\phi}S_g &= \frac{1}{2\kappa^2}\int_Mh\left[(f_{\phi} + Z_{\phi}g^{\mu\nu}\phi_{,\mu}\phi_{,\nu})\delta\phi + 2Zg^{\mu\nu}\phi_{,\nu}\delta\phi_{,\mu}\right]d^4x\\
&= \frac{1}{2\kappa^2}\int_Mh\left[f_{\phi} + Z_{\phi}g^{\mu\nu}\phi_{,\mu}\phi_{,\nu} - 2g^{\mu\nu}\lc{\nabla}_{,\mu}(Z\phi_{,\nu})\right]\delta\phi\,d^4x\,,
\end{split}
\end{equation}
where we have performed integration by parts to arrive at the second line. The corresponding variation of the matter part of the action vanishes, \(\delta_{\phi}S_m = 0\), since we do not consider any direct coupling between the scalar field and matter fields. Hence, the field equation does not contain a source term, and can finally be brought into the form
\begin{equation}\label{eqn:scalarfeq}
f_{\phi} - Z_{\phi}g^{\mu\nu}\phi_{,\mu}\phi_{,\nu} - 2Z\lc{\square}\phi = 0\,,
\end{equation}
where \(\lc{\square} = g^{\mu\nu}\lc{\nabla}_{\mu}\lc{\nabla}_{\nu}\) is the d'Alembert operator.

Let us note that contrary to the scalar-curvature theories the second derivatives of the tetrad (or metric) do not appear in the scalar field equation, and thus the procedure of ``debraiding'' (c.f.~\cite{Bettoni:2015wta}) is not necessary. As a consequence, when the matter Lagrangian does not explicitly involve the scalar field, the scalar field equation remains without a matter contribution as a source term, e.g., the trace of the matter energy-momentum tensor. Therefore in scalar-torsion gravity the chameleon screening mechanism does not work as in the scalar-curvature theories \cite{Khoury:2003aq,Burrage:2017qrf}, unless one introduces a coupling of the scalar field to matter~\cite{Sadjadi:2015fca} or a boundary term~\cite{Bamba:2013jqa,Bahamonde:2015hza,Bahamonde:2017wwk}.

\subsection{Relation between field equations}\label{ssec:feqrel}
It has recently been shown for \(f(T)\) gravity and more general teleparallel gravity theories with second order field equations that the antisymmetric part of the tetrad field equations is identical to the connection field equations, so that the flat spin connection is a pure gauge degree of freedom corresponding to the local Lorentz invariance of the action~\cite{Golovnev:2017dox,Krssak:2017nlv,BeltranJimenez:2017tkd,Hohmann:2017duq}. We now show that the same holds true also for the class of scalar-torsion theories presented here. For this purpose we consider the antisymmetric part of the tetrad field equations~\eqref{eqn:tetradfeq}, which reads
\begin{equation}\label{eqn:asymder}
\begin{split}
0 &= 2\lc{\nabla}_{\rho}\left(f_TS_{[\nu\mu]}{}^{\rho}\right) + f_T\left(T^{\rho}{}_{\rho\sigma}T^{\sigma}{}_{\mu\nu} + T^{\rho\sigma}{}_{[\nu}T_{\mu]\rho\sigma}\right)\\
&= 3\partial_{[\rho}f_TT^{\rho}{}_{\mu\nu]} + f_T\left(3\lc{\nabla}_{[\rho}T^{\rho}{}_{\mu\nu]} + T^{\rho}{}_{\rho\sigma}T^{\sigma}{}_{\mu\nu} - T^{\rho\sigma}{}_{[\mu}T_{\nu]\rho\sigma}\right)\,,
\end{split}
\end{equation}
where we have used the definitions~\eqref{eqn:suppot} of the superpotential and~\eqref{eqn:contor} of the contortion.
 For the first term in brackets we now expand the covariant derivative into Christoffel symbols,
\begin{equation}
\lc{\nabla}_{[\rho}T^{\rho}{}_{\mu\nu]} = \partial_{[\rho}T^{\rho}{}_{\mu\nu]} + \lc{\Gamma}^{\rho}{}_{\sigma[\rho}T^{\sigma}{}_{\mu\nu]} - \lc{\Gamma}^{\sigma}{}_{[\mu\rho}T^{\rho}{}_{|\sigma|\nu]} - \lc{\Gamma}^{\sigma}{}_{[\nu\rho}T^{\rho}{}_{\mu]\sigma}\,,\\
\end{equation}
where the last two terms vanish due to the symmetry of the Christoffel symbols in their lower indices, since the Levi-Civita connection is torsion free. For the remaining term we express the Levi-Civita connection through the teleparallel connection and the contortion,
\begin{equation}
\lc{\Gamma}^{\rho}{}_{\sigma[\rho}T^{\sigma}{}_{\mu\nu]} = \tp{\Gamma}^{\rho}{}_{\sigma[\rho}T^{\sigma}{}_{\mu\nu]} - K^{\rho}{}_{\sigma[\rho}T^{\sigma}{}_{\mu\nu]}\,.
\end{equation}
A direct calculation then shows that
\begin{equation}
\partial_{[\rho}T^{\rho}{}_{\mu\nu]} + \tp{\Gamma}^{\rho}{}_{\sigma[\rho}T^{\sigma}{}_{\mu\nu]} = \tp{R}^{\rho}{}_{[\mu\nu]\rho} = 0\,,
\end{equation}
which vanishes, since the teleparallel spin connection, and hence its spacetime connection are flat,
\begin{equation}
\tp{R}^{\rho}{}_{\sigma\mu\nu} = h_a{}^{\rho}h^b{}_{\sigma}\tp{R}^a{}_{b\mu\nu} = 0\,.
\end{equation}
We are thus left with the only remaining term, which reads
\begin{equation}\label{eqn:dttott}
\lc{\nabla}_{[\rho}T^{\rho}{}_{\mu\nu]} = -K^{\rho}{}_{\sigma[\rho}T^{\sigma}{}_{\mu\nu]} = -\frac{1}{3}\left(T^{\rho}{}_{\mu\nu}T^{\sigma}{}_{\sigma\rho} - T^{\rho\sigma}{}_{[\mu}T_{\nu]\rho\sigma}\right)\,.
\end{equation}
Hence, the bracket in the second line of the antisymmetric field equation~\eqref{eqn:asymder} vanishes. This equation thus reduces to
\begin{equation}\label{eqn:asymfeq}
\partial_{[\rho}f_TT^{\rho}{}_{\mu\nu]} = 0\,,
\end{equation}
which indeed agrees with the connection field equations~\eqref{eqn:connfeq}.

We are finally left with the symmetric part of the field equations
\begin{multline}
\frac{1}{2}fg_{\mu\nu} + \lc{\nabla}_{\rho}\left(f_TS_{(\mu\nu)}{}^{\rho}\right) + \frac{1}{2}f_T\left(2T^{\rho}{}_{\rho\sigma}T_{(\mu\nu)}{}^{\sigma} - \frac{1}{2}T_{\mu\rho\sigma}T_{\nu}{}^{\rho\sigma} + T^{\rho\sigma}{}_{(\mu}T_{\nu)\rho\sigma}\right)\\
- Z\phi_{,\mu}\phi_{,\nu} + \frac{1}{2}Zg_{\mu\nu}g^{\rho\sigma}\phi_{,\rho}\phi_{,\sigma} = \kappa^2\Theta_{\mu\nu}\,.
\end{multline}
Using the relation
\begin{equation}
S_{(\mu}{}^{\rho\sigma}T_{\nu)\rho\sigma} = -2T^{\rho}{}_{\rho\sigma}T_{(\mu\nu)}{}^{\sigma} + \frac{1}{2}T_{\mu\rho\sigma}T_{\nu}{}^{\rho\sigma} - T^{\rho\sigma}{}_{(\mu}T_{\nu)\rho\sigma}
\end{equation}
it can also be written as
\begin{equation}\label{eqn:symfeq}
\frac{1}{2}fg_{\mu\nu} + \lc{\nabla}_{\rho}\left(f_TS_{(\mu\nu)}{}^{\rho}\right) - \frac{1}{2}f_TS_{(\mu}{}^{\rho\sigma}T_{\nu)\rho\sigma} - Z\phi_{,\mu}\phi_{,\nu} + \frac{1}{2}Zg_{\mu\nu}g^{\rho\sigma}\phi_{,\rho}\phi_{,\sigma} = \kappa^2\Theta_{\mu\nu}\,.
\end{equation}
These equations remain to be solved independently from the antisymmetric part~\eqref{eqn:asymfeq}.

To recap, the dynamical equations for the theory \eqref{eqn:action} are the ten symmetric equations \eqref{eqn:symfeq} for the tetrad components, the scalar field equation \eqref{eqn:scalarfeq}, and the matter equations of motion (not specified here). Demanding flatness constrains the spin connection so that only six components (or combinations of components) remain free. These six freedoms in the spin connection get fixed by the conditions \eqref{eqn:connection condition} (or in an equivalent form \eqref{eqn:asymfeq}).
The ten dynamical tetrad components match the number of independent metric components and describe gravity, while the other six tetrad components characterize the frame of the local observer. Choosing a particular local observer fixes these six tetrad components which in turn completely fixes the spin connection, the latter encoding the inertial effects in the observer frame \cite{Krssak:2017nlv}.

\subsection{Conservation of matter energy-momentum}\label{ssec:enmomcons}
We finally show that the covariant conservation of the energy-momentum energy-momentum tensor can also be derived from the gravitational field equations. For this purpose, we take the covariant divergence of the symmetric field equation~\eqref{eqn:symfeq}, which reads
\begin{equation}\label{eqn:telebianchi}
\begin{split}
\kappa^2\lc{\nabla}^{\mu}\Theta_{\mu\nu} &= -Z'g^{\mu\rho}\phi_{,\mu}\phi_{,\rho}\phi_{,\nu} - Z\lc{\square}\phi\phi_{,\nu} - Z\phi_{,\mu}\lc{\nabla}^{\mu}\lc{\nabla}_{\nu}\phi + \frac{1}{2}Z'g^{\rho\sigma}\phi_{,\rho}\phi_{,\sigma}\phi_{,\nu} + Z\phi_{,\rho}\lc{\nabla}_{\nu}\lc{\nabla}^{\rho}\phi + \frac{1}{2}f_{\phi}\phi_{,\nu}\\
&\phantom{=}+ f_T\lc{\nabla}^{\mu}\left(\lc{\nabla}_{\rho}S_{(\mu\nu)}{}^{\rho} - \frac{1}{2}S_{(\mu}{}^{\rho\sigma}T_{\nu)\rho\sigma} + \frac{1}{2}Tg_{\mu\nu}\right)\\
&\phantom{=}+ \lc{\nabla}^{\mu}\lc{\nabla}_{\rho}f_TS_{(\mu\nu)}{}^{\rho} - \lc{\nabla}^{\mu}f_T\left(\frac{1}{2}S_{(\mu}{}^{\rho\sigma}T_{\nu)\rho\sigma} - \lc{\nabla}^{\rho}S_{(\mu\nu)\rho} - \lc{\nabla}^{\rho}S_{(\rho\nu)\mu}\right)\,.
\end{split}
\end{equation}
In the first line two terms cancel since the Levi-Civita connection is torsion free, which implies
\begin{equation}\label{eqn:notorsion}
\lc{\nabla}_{\mu}\lc{\nabla}_{\nu}\psi - \lc{\nabla}_{\nu}\lc{\nabla}_{\mu}\psi = 0
\end{equation}
for any scalar function \(\psi\), in particular also for \(\psi = \phi\). The remaining terms in the first line can then be written as
\begin{equation}
\frac{1}{2}\phi_{,\nu}\left(f_{\phi} - Z'g^{\mu\rho}\phi_{,\mu}\phi_{,\rho} - 2Z\lc{\square}\phi\right)\,,
\end{equation}
and one recognizes that the term in brackets is simply the left hand side of the scalar field equation~\eqref{eqn:scalarfeq}, and thus vanishes. For the second line one once again makes use of the geometric identity~\eqref{eqn:stor} to realize that the term in brackets is simply the Einstein tensor; its covariant divergence vanishes due to the Bianchi identity. We are left with the third line. For its first term one finds the identity
\begin{equation}\label{eqn:dasymfeq1}
\lc{\nabla}^{\mu}\lc{\nabla}_{\rho}f_TS_{(\mu\nu)}{}^{\rho} = \frac{1}{2}\left(\lc{\square}f_TT^{\rho}{}_{\rho\mu} - \lc{\nabla}_{\mu}\lc{\nabla}_{\rho}f_TT^{\mu\rho}{}_{\nu} - \lc{\nabla}_{\mu}\lc{\nabla}_{\nu}f_TT_{\rho}{}^{\rho\mu}\right) = -\frac{3}{2}\lc{\nabla}^{\mu}\lc{\nabla}_{[\rho}f_TT^{\rho}{}_{\mu\nu]}\,,
\end{equation}
where we have once more used the symmetry~\eqref{eqn:notorsion} of the Levi-Civita connection, now with \(\psi = f_T\). For the last two terms in the third line we find
\begin{equation}
\lc{\nabla}^{\rho}S_{(\mu\nu)\rho} + \lc{\nabla}^{\rho}S_{(\rho\nu)\mu} = -\frac{3}{2}\lc{\nabla}_{[\rho}T^{\rho}{}_{\mu\nu]} + \frac{1}{2}\left(\lc{\nabla}_{\rho}T_{\mu\nu}{}^{\rho} + \lc{\nabla}_{\mu}T^{\rho}{}_{\rho\nu} - g_{\mu\nu}\lc{\nabla}_{\rho}T_{\sigma}{}^{\sigma\rho}\right)\,.
\end{equation}
Contracting the terms in brackets with \(\lc{\nabla}^{\mu}f_T\) we obtain
\begin{equation}\label{eqn:dasymfeq2}
\frac{1}{2}\lc{\nabla}^{\mu}f_T\left(\lc{\nabla}_{\rho}T_{\mu\nu}{}^{\rho} + \lc{\nabla}_{\mu}T^{\rho}{}_{\rho\nu} - g_{\mu\nu}\lc{\nabla}_{\rho}T_{\sigma}{}^{\sigma\rho}\right) = -\frac{3}{2}\lc{\nabla}_{[\rho}f_T\lc{\nabla}^{\mu}T^{\rho}{}_{\mu\nu]}\,.
\end{equation}
Now the terms~\eqref{eqn:dasymfeq1} and~\eqref{eqn:dasymfeq2} are combined to the covariant divergence of the connection field equation~\eqref{eqn:connfeq}, and hence vanish. Using the identity~\eqref{eqn:dttott}, the energy-momentum conservation law~\eqref{eqn:telebianchi} reduces to
\begin{equation}
\kappa^2\lc{\nabla}^{\mu}\Theta_{\mu\nu} = -\frac{1}{2}\lc{\nabla}^{\mu}f_T\left(S_{(\mu}{}^{\rho\sigma}T_{\nu)\rho\sigma} - T^{\rho}{}_{\mu\nu}T^{\sigma}{}_{\sigma\rho} + T^{\rho\sigma}{}_{[\mu}T_{\nu]\rho\sigma}\right)\,.
\end{equation}
It is helpful to realize that the term in brackets can be written as
\begin{equation}
-\frac{1}{2}\left(S_{(\mu}{}^{\rho\sigma}T_{\nu)\rho\sigma} - T^{\rho}{}_{\mu\nu}T^{\sigma}{}_{\sigma\rho} + T^{\rho\sigma}{}_{[\mu}T_{\nu]\rho\sigma}\right) = S_{\rho\sigma\mu}K^{\rho\sigma}{}_{\nu}\,.
\end{equation}
Note that the contortion tensor~\eqref{eqn:contor} is antisymmetric in its first two indices. Hence, only the part of the superpotential contributes which is likewise antisymmetric in its first two indices. Finally, we find that
\begin{equation}
\lc{\nabla}_{\mu}f_TS_{[\rho\sigma]}{}^{\mu} = -\frac{3}{2}\lc{\nabla}_{[\mu}f_TT^{\mu}{}_{\rho\sigma]}\,,
\end{equation}
which vanishes due to the connection field equations~\eqref{eqn:connfeq}. Hence, the right hand side of the conservation equations~\eqref{eqn:telebianchi} vanishes, as one would expect. Let us note that the spin connection equation played a role in providing that.

\subsection{Reduction to general relativity}\label{ssec:grred}
The field equations~\eqref{eqn:asymfeq} and~\eqref{eqn:symfeq} have the interesting property that for particular solutions they reduce to the field equations of general relativity (GR). These solutions must satisfy the conditions that both \(T\) and \(\phi\) are constant, and \(f_{\phi}(T, \phi) = 0\) for these constant values. Note that constant torsion scalar does not require constant tetrad and spin connection, there are field configurations where those variables cancel each other in the torsion scalar. If both \(T\) and \(\phi\) are constant with respect to spacetime, i.e., their derivatives with respect to spacetime directions vanish, the same holds for any function of these variables, and thus in particular for \(f\) and its derivatives. As an immediate consequence, the antisymmetric part~\eqref{eqn:asymfeq} of the field equations is solved identically. Further, the scalar field equation~\eqref{eqn:scalarfeq} reduces to \(f_{\phi} = 0\), and is solved due to our assumptions. It remains to show that the symmetric part~\eqref{eqn:symfeq} reduces to the general relativity field equations. For this purpose, we write these equations in the form
\begin{equation}
\frac{1}{2}fg_{\mu\nu} + f_T\left(\lc{\nabla}_{\rho}S_{(\mu\nu)}{}^{\rho} - \frac{1}{2}S_{(\mu}{}^{\rho\sigma}T_{\nu)\rho\sigma}\right) = \kappa^2\Theta_{\mu\nu}\,,
\end{equation}
where we used the constancy of \(\phi\) to omit the scalar field kinetic terms, and the constancy of \(f_T\) to remove its contribution to the derivative in the second term. We can now split the first term to obtain
\begin{equation}
\frac{1}{2}\left(f - f_TT\right)g_{\mu\nu} + f_T\left(\lc{\nabla}_{\rho}S_{(\mu\nu)}{}^{\rho} - \frac{1}{2}S_{(\mu}{}^{\rho\sigma}T_{\nu)\rho\sigma} + \frac{1}{2}Tg_{\mu\nu}\right) = \kappa^2\Theta_{\mu\nu}\,.
\end{equation}
One further uses the identity
\begin{equation}\label{eqn:stor}
\lc{\nabla}_{\rho}S_{(\mu\nu)}{}^{\rho} - \frac{1}{2}S_{(\mu}{}^{\rho\sigma}T_{\nu)\rho\sigma} + \frac{1}{2}Tg_{\mu\nu} = \lc{R}_{\mu\nu} - \frac{1}{2}\lc{R}g_{\mu\nu}\,,
\end{equation}
so that the field equations finally take the form
\begin{equation}
\frac{1}{2}\left(f - f_TT\right)g_{\mu\nu} + f_T\left(\lc{R}_{\mu\nu} - \frac{1}{2}\lc{R}g_{\mu\nu}\right) = \kappa^2\Theta_{\mu\nu}\,.
\end{equation}
Note that the coefficients of the metric and the Einstein tensor on the left hand side are both constant with respect to spacetime, due to our assumptions, and can thus be related to effective cosmological and gravitational constants. It follows that the metric~\eqref{eqn:metric} satisfies the Einstein equations corresponding to these two parameters.

\section{Determining the spin connection}\label{sec:connection}

After deriving the field equations, the aim of this section is to analyze the connection equation~\eqref{eqn:connection condition} and provide a sample of tetrads, spin connections and scalar fields which solve it, and thus can serve as a starting point for finding solutions to the remaining field equations~\eqref{eqn:symfeq} and~\eqref{eqn:scalarfeq}. We start with a few general considerations in section~\ref{ssec:genconn}, before providing a number of specific examples. In particular, we discuss Friedmann-Lema\^itre-Robertson-Walker spacetimes in section~\ref{ssec:flrw}, Kasner spacetimes in section~\ref{ssec:kasner} and spherically symmetric spacetimes in section~\ref{ssec:spherical}.

\subsection{General considerations}\label{ssec:genconn}

In principle there is a reasonable way how to approach the set of field equations. First one should take an ansatz for the tetrad (which may implicitly relate to a certain Lorentz observer) as well as for the scalar field. Second, demanding that the curvature \eqref{eqn:curv} vanishes, and the conditions \eqref{eqn:connection condition} are satisfied when the ansatz is substituted in, should determine the form of the spin connection associated with that ansatz. Then the tetrad ansatz and the spin connection can be substituted into the tetrad field equations  \eqref{eqn:symfeq} and the scalar field equations \eqref{eqn:scalarfeq} to be solved. Finally one may use local Lorentz transformations \eqref{eqn:Lorentz transformation} to find other equivalent forms of the
solution (corresponding to different local observers).

Let us look more closely at the conditions on the spin connection \eqref{eqn:connection condition} or \eqref{eqn:asymfeq}. We may encounter several different situations, some of those have been noted before for $f(T)$ gravity \cite{Krssak:2017nlv}, and have actually come up even before while trying to satisfy the antisymmetric part of the tetrad field equations \cite{Boehmer:2011gw,Tamanini:2012hg}.
First, if $f_T$ is constant, i.e., \(f_{TT} = f_{T\phi} \equiv 0\), the constraint is identically satisfied. This case pertains to taking the theory to be the teleparallel equivalent of general relativity (with the optional scalar field minimally coupled). There the flat spin connection can be chosen completely arbitrarily, but there can be other principles to constrain it besides the field equations \cite{Krssak:2015rqa,Krssak:2015lba}.

Second, we may choose the spin connection such that  $\partial_\mu T=0$ and $\partial_\mu \phi=0$. This is related to a possible strategy of looking for solutions in $f(T)$ gravity where one tries to Lorentz transform the ansatz tetrad of interest into a frame where the torsion scalar $T$ or its derivative vanishes \cite{Ferraro:2011ks,Boehmer:2011gw,Tamanini:2012hg,Bejarano:2014bca,Bejarano:2017akj}. The strategy is good, since in that particular frame the connection condition is automatically satisfied for any $\tp{\omega}^a{}_{b\mu}$, including also a zero spin connection. Therefore omitting the spin connection in such a frame while solving the tetrad field equations is a consistent move. However, the drawback is that the tetrad field equations reduce to those of teleparallel equivalent of general relativity, as we have briefly shown in section~\ref{ssec:grred}. Hence with this method one only recovers the solutions already present in general relativity. But this method is still a nice way to learn about universal solutions, i.e., solutions which are common to the whole $f(T,\phi)$ family of theories.

Third, due to the properties of the ansatz it might be possible to solve the spin connection condition independently of the function $f$. This can happen when $\partial_\mu T$ and $\partial_\mu \phi$ summed over with the terms in the brackets in Eq.~\eqref{eqn:connection condition} yield zero. This is a much more interesting option, since the whole set of equations is not necessarily reduced to the teleparallel equivalent of general relativity, for the tetrad and scalar field equations still involve the function $f$. Thus these solutions may extend the repertoire of general relativity. A feasible way to realize this situation is when the ansatz depends on one particular coordinate \(y\), like time \(y = t\) in cosmology or radial distance \(y = r\) in static, spherically symmetric spacetime. Then the two scalars \(T\) and \(\phi\), and hence \(f_T\), depend only on this coordinate. As a consequence, \(\partial_{\mu}f_T \propto \partial_{\mu}y\), and the particular choice of the function \(f\) becomes irrelevant for solving the connection field equations. From a geometric point of view, the expression in the brackets in Eq.~\eqref{eqn:connection condition} can be interpreted as a set of six vectors labeled by the six possible values of the antisymmetric index pair \([ab]\), and the equations are solved if these vectors are tangent to the hypersurfaces of constant \(y\). In the following subsections we illustrate how this works in a few examples. In principle it may happen that the situation can be realized with more general ans\"atze as well, due to some underlying symmetry.

Finally, in the more general case, it might not be possible to solve the spin connection equation independently of the function $f$. In the computationally worst case one may have to tackle all equations  \eqref{eqn:connection condition}, \eqref{eqn:symfeq}, \eqref{eqn:scalarfeq} simultaneously. With nonminimally coupled scalar field there might also exist some particularly amenable forms of the function $f(T,\phi)$ which could allow cancellation in \eqref{eqn:connection condition f_T written out} and thus turn out to be helpful in finding the solutions. Note that also in this situation the aforementioned geometric interpretation holds, but the hypersurfaces are defined by constant values of \(f_T\), and thus depend on the choice of the function \(f\).

In the following subsections we present some tetrads in their diagonal form together with a non-vanishing spin connection which satisfies the condition \eqref{eqn:connection condition}.

% ; an equivalent description in the Weitzenböck gauge uses non-diagonal tetrad and a vanishing spin connection~\cite{HJKP:2018sym}.

\subsection{Friedmann-Lema\^itre-Robertson-Walker spacetimes}\label{ssec:flrw}

Homogeneous and isotropic cosmological spacetimes are described by the metric in the Friedmann-Lema\^itre-Robertson-Walker (FLRW) form
\begin{equation}
\label{eqn:flrwmetric}
g_{\mu\nu}dx^{\mu}dx^{\nu} = dt^2 - a(t)\left[\frac{dr^2}{1 - kr^2} + r^2(d\vartheta^2 + \sin^2\vartheta d\varphi^2)\right] \,,
\end{equation}
written in spherical coordinates \(t,r,\vartheta,\varphi\). Here \(k \in \{-1,0,1\}\) determines the sign of the spatial curvature. Note that in all three cases the cosmological symmetry imposes that the scalar field is evolving homogeneously, $\phi = \phi(t)$, while the matter energy-momentum tensor must be given by an ideal fluid,
\begin{equation}
\Theta^{\mu\nu} = (\rho + p)u^{\mu}u^{\nu} - pg^{\mu\nu}
\end{equation}
with energy density \(\rho = \rho(t)\), pressure \(p = p(t)\) and four-velocity \(u^{\mu} = \partial_t\) normalized by the metric, \(g_{\mu\nu}u^{\mu}u^{\nu} = 1\). A canonical choice for the tetrad is the diagonal one,
\begin{equation}
\label{eqn:flrwtetrad}
h^a{}_{\mu} = \mathrm{diag}\left(1, \frac{a(t)}{\sqrt{1 - kr^2}}, a(t)r, a(t)r\sin\vartheta\right)\,.
\end{equation}
Following the procedure detailed in the general discussion, one then determines the spin connection by solving the antisymmetric part~\eqref{eqn:connfeq} of the field equations. Depending on the value of \(k\), one may find different solutions:

\begin{itemize}
\item
In the case \(k = 0\) one may use the spin connection~\cite{Krssak:2015oua}
\begin{equation}
\label{eqn:k=0 spherical connection}
\tp{\omega}^{1}{}_{2\vartheta} = -\tp{\omega}^{2}{}_{1\vartheta} = -1 \,, \qquad
\tp{\omega}^{1}{}_{3\varphi} = -\tp{\omega}^{3}{}_{1\varphi} = -\sin\vartheta \,, \qquad
\tp{\omega}^{2}{}_{3\varphi} = -\tp{\omega}^{3}{}_{2\varphi} = -\cos\vartheta \,.
\end{equation}
One then finds that the remaining tetrad field equations are given by
\begin{eqnarray}
\frac{1}{2}f + 6f_TH^2 - \frac{1}{2}Z\dot{\phi}^2 &=& \kappa^2\rho\,,\\
\frac{1}{2}f + 2f_{T\phi}H\dot{\phi} - 24f_{TT}\dot{H}H^2 + 6f_TH^2 + 2f_T\dot{H} + \frac{1}{2}Z\dot{\phi}^2 &=& -\kappa^2p\,,
\end{eqnarray}
where $H=\tfrac{\dot{a}}{a}$ is the Hubble parameter, and a dot denotes the derivative with respect to \(t\). Note that Minkowski spacetime is included as the special case \(a = 1\).

\item
For \(k = 1\), a viable spin connection is given by~\cite{HJKP:2018sym}
\begin{gather}
\tp{\omega}^1{}_{2\vartheta} = -\tp{\omega}^2{}_{1\vartheta} = -\sqrt{1 - r^2}\,, \qquad
\tp{\omega}^1{}_{2\varphi} = -\tp{\omega}^2{}_{1\varphi} = -r\sin\vartheta\,, \qquad
\tp{\omega}^1{}_{3\vartheta} = -\tp{\omega}^3{}_{1\vartheta} = r\,, \nonumber\\
\tp{\omega}^1{}_{3\varphi} = -\tp{\omega}^3{}_{1\varphi} = -\sqrt{1 - r^2}\sin\vartheta\,, \qquad
\tp{\omega}^2{}_{3r} = -\tp{\omega}^3{}_{2r} = -\frac{1}{\sqrt{1 - r^2}}\,, \qquad
\tp{\omega}^2{}_{3\varphi} = -\tp{\omega}^3{}_{2\varphi} = -\cos\vartheta\,.
\label{eqn:k=+1 spin connection}
\end{gather}
In this case one obtains the remaining tetrad field equations
\begin{eqnarray}
\frac{1}{2}f + 6f_TH^2 - \frac{1}{2}Z\dot{\phi}^2 &=& \kappa^2\rho\,,\\
\frac{1}{2}f + 2f_{T\phi}H\dot{\phi} - 24f_{TT}\left(\dot{H} + \frac{1}{a^2}\right)H^2 + 6f_TH^2 + 2f_T\left(\dot{H} - \frac{1}{a^2}\right) + \frac{1}{2}Z\dot{\phi}^2 &=& -\kappa^2p\,.
\end{eqnarray}

\item
Finally, for \(k = -1\) one may use the spin connection~\cite{HJKP:2018sym}
\begin{gather}
\tp{\omega}^0{}_{1r} = \tp{\omega}^1{}_{0r} = \frac{1}{\sqrt{1 + r^2}}\,, \qquad
\tp{\omega}^0{}_{2\vartheta} = \tp{\omega}^2{}_{0\vartheta} = r\,, \qquad
\tp{\omega}^0{}_{3\varphi} = \tp{\omega}^3{}_{0\varphi} = r\sin\vartheta\,, \nonumber
\\
\tp{\omega}^1{}_{2\vartheta} = -\tp{\omega}^2{}_{1\vartheta} = -\sqrt{1 + r^2}\,, \qquad
\tp{\omega}^1{}_{3\varphi} = -\tp{\omega}^3{}_{1\varphi} = -\sqrt{1 + r^2}\sin\vartheta\,, \qquad
\tp{\omega}^2{}_{3\varphi} = -\tp{\omega}^3{}_{2\varphi} = -\cos\vartheta\,.
\label{eqn:k=-1 spin connection}
\end{gather}
Here the remaining tetrad field equations are given by
\begin{eqnarray}
\frac{1}{2}f + 6f_TH\left(H - \frac{1}{a}\right) - \frac{1}{2}Z\dot{\phi}^2 &=& \kappa^2\rho\,,\\
\hspace*{-12mm}\frac{1}{2}f + 2f_{T\phi}\left(H - \frac{1}{a}\right)\dot{\phi} - 24f_{TT}\left(\dot{H} + \frac{H}{a}\right)\left(H - \frac{1}{a}\right)^2 + 6f_TH\left(H - \frac{1}{a}\right) + 2f_T\left(\dot{H} + \frac{1}{a^2}\right) + \frac{1}{2}Z\dot{\phi}^2 &=& -\kappa^2p\,.
\end{eqnarray}
\end{itemize}

In all three cases the scalar field equation reduces to
\begin{equation}
f_{\phi} - 2Z\ddot{\phi} - 6ZH\dot{\phi} - Z_{\phi}\dot{\phi}^2 = 0\,.
\end{equation}
One easily checks that the three mentioned spin connections satisfy the condition~\eqref{eqn:connection condition} for the connection, due to the reasons outlined as the third option in subsec. \ref{ssec:genconn}. Let us emphasize that the tetrad \eqref{eqn:flrwtetrad} alone with a vanishing spin connection does not solve the condition~\eqref{eqn:connection condition}. It is worth noting that the structure of the spin connection components as well as the tetrad field equations in the $k=+1$ \eqref{eqn:k=+1 spin connection} and $k=-1$ \eqref{eqn:k=-1 spin connection} cases is different. For example in the latter case there are nontrivial time components which do not appear in the former case. In the $k=1$ case it was found that a certain local Lorentz rotation can make a transformation into a local frame, where all the spin connection components vanish, but the tetrad components become nondiagonal and more complicated \cite{Tamanini:2012hg}. In the $k=-1$ case, the same can happen, but one needs to employ a Lorentz boost \cite{HJKP:2018sym}. The remaining tetrad and scalar field equations we displayed above are invariant under this simultaneous transformation of the tetrad and the spin connection.

\subsection{Kasner spacetimes}\label{ssec:kasner}

Homogeneous, but in general anisotropic Kasner spacetime can be realized by choosing the tetrad
\begin{equation}\label{eqn:kasnertetrad}
h^a_{\ \mu} = \mathrm{diag}(1, a(t), b(t), c(t)),
\end{equation}
in Cartesian coordinates. This, together with a vanishing spin connection,
\begin{equation}\label{eqn:zeroconn}
\tp{\omega}^a{}_{b\mu} = 0\,,
\end{equation}
makes the connection condition~\eqref{eqn:connection condition}  satisfied, which vindicates the respective studies in $f(T)$ gravity \cite{Rodrigues:2012qua,Paliathanasis:2016vsw,Paliathanasis:2017htk,Skugoreva:2017vde}. But for more general anisotropic models one has to check the compatibility of the spin connection.

In the special case \(a = b = c\) the tetrad~\eqref{eqn:kasnertetrad} reduces to another viable tetrad to describe FLRW spacetime with \(k = 0\),
\begin{equation}
h^a_{\ \mu} = \mathrm{diag}(1, a(t), a(t), a(t)) \,.
\end{equation}
Along with a vanishing spin connection~\eqref{eqn:zeroconn} one finds that the condition~\eqref{eqn:connection condition} is easily satisfied. Since here assuming the vanishing spin connection yields the correct result, the former studies of the spatially flat cosmology in $f(T)$ and scalar-torsion gravity remain valid, see e.g.\  Refs.~\cite{Farrugia:2016qqe,Capozziello:2017bxm,Mirza:2017vrk,Hohmann:2017jao,Capozziello:2017uam,Chakrabarti:2017moe} for a selection of latest works.

\subsection{Spherically symmetric spacetimes}\label{ssec:spherical}

A general static spherically symmetric spacetime can be represented by a diagonal tetrad
\begin{equation} \label{eqn:spherical tetrad}
h^a_{\ \mu} = \mathrm{diag}(A(r), B(r),r, r\sin \vartheta) \,.
\end{equation}
This, together with the flat spin connection \cite{Krssak:2015oua}
\begin{equation}
\tp{\omega}^{1}{}_{2\vartheta} = -\tp{\omega}^{2}{}_{1\vartheta} = -1 \,, \qquad
\tp{\omega}^{1}{}_{3\varphi} = -\tp{\omega}^{3}{}_{1\varphi} = -\sin\vartheta \,, \qquad
\tp{\omega}^{2}{}_{3\varphi} = -\tp{\omega}^{3}{}_{2\varphi} = -\cos\vartheta \,
\label{eqn:spherical spin connection}
\end{equation}
and a generic spherically symmetric ansatz for the scalar field, $\phi=\phi(r)$, satisfies the condition for the connection \eqref{eqn:connection condition}. As with the previous cases presented in spherical coordinates, it is possible by a Lorentz transformation to find a frame where the spin connection vanishes, while the tetrad becomes nondiagonal \cite{Boehmer:2011gw}. Thus the earlier research on the spherically symmetric systems in $f(T)$ gravity based on this nondiagonal tetrad  \cite{Boehmer:2011gw,Tamanini:2012hg,Horvat:2014xwa,Iorio:2015rla,Ruggiero:2015oka,Ruggiero:2016iaq,Farrugia:2016xcw} has a valid starting point, albeit sees physics in a particular frame. Assuming the diagonal tetrad \eqref{eqn:spherical tetrad} must happen in conjunction with the nonzero spin connection \eqref{eqn:spherical spin connection}, like in Ref.~\cite{DeBenedictis:2016aze}, otherwise one is led to wrong results.

\section{Generalization to multiple scalar fields}\label{sec:multi}
In the previous sections of this article we have considered a class of scalar-torsion theories of gravity obtained by (in general nonminimally) coupling a scalar field to teleparallel gravity. In this section we extend our findings to multiple scalar fields. For this purpose we consider a multiplet \(\boldsymbol{\phi} = (\phi^A), A = 1, \ldots, N\) of \(N\) scalar fields. We then replace the action~\eqref{eqn:action} by the more general form
\begin{equation}\label{eqn:mulaction}
S = \frac{1}{2\kappa^2}\int_M\left[f(T,\boldsymbol{\phi}) + Z_{AB}(\boldsymbol{\phi})g^{\mu\nu}\phi^A_{,\mu}\phi^B_{,\nu}\right]\theta d^4x + S_m[\theta^a, \chi^I]\,.
\end{equation}
This action differs from the single field case in two ways. First, in order to potentially render all scalar fields dynamical, there must be a kinetic term involving all fields. A natural generalization of the single scalar field kinetic term is to equip the parameter function \(Z\) with two scalar field indices, such that it is symmetric in these indices, \(Z_{[AB]} = 0\); any antisymmetric part would cancel due to the contraction with a symmetric tensor composed from the derivatives of the scalar fields. Second, the two parameter functions \(f\) and \(Z_{AB}\) now depend on all scalar fields, and hence on the field multiplet \(\boldsymbol{\phi}\). This must be taken into account when calculating variations and derivatives of the parameter functions. (In special cases like $f(T,\boldsymbol{\phi})=f(\boldsymbol{\phi})T$, it is possible to redefine the scalar fields so that only one of them is nonminimally coupled, as pointed out in the multiscalar-curvature theory \cite{Hohmann:2016yfd}, but we will not delve into this option here.)

It is now straightforward to derive the field equations from the action~\eqref{eqn:mulaction}, essentially following the same steps as shown explicitly in section~\ref{sec:feqs}. We will not repeat these steps here, and only display the field equations in their final form. We start with the symmetric part~\eqref{eqn:symfeq}, which generalizes to
\begin{equation}\label{eqn:mulsymfeq}
\frac{1}{2}fg_{\mu\nu} + \lc{\nabla}_{\rho}\left(f_TS_{(\mu\nu)}{}^{\rho}\right) - \frac{1}{2}f_TS_{(\mu}{}^{\rho\sigma}T_{\nu)\rho\sigma} - Z_{AB}\phi^A_{,\mu}\phi^B_{,\nu} + \frac{1}{2}Z_{AB}\phi^A_{,\rho}\phi^B_{,\sigma}g^{\rho\sigma}g_{\mu\nu} = \kappa^2\Theta_{\mu\nu}\,,
\end{equation}
where only the terms originating from the kinetic energy of the scalar fields are visibly affected. These do not appear in the antisymmetric part~\eqref{eqn:asymfeq} of the tetrad field equations, which hence retain their form,
\begin{equation}\label{eqn:mulasymfeq}
\partial_{[\rho}f_TT^{\rho}{}_{\mu\nu]} = 0\,.
\end{equation}
This in particular implies that the connections given in section~\ref{sec:connection} remain valid also in the case of multiple scalar fields.

Finally, the generalization of the scalar field equation~\eqref{eqn:scalarfeq} requires more attention, since derivatives of \(Z_{AB}\) now carry different types of indices, and the correct indices must be chosen in contractions. Starting from the action~\eqref{eqn:mulaction} we find the scalar field equations
\begin{equation}\label{eqn:mulscalarfeq}
f_{\phi^A} - \left(2Z_{AB,\phi^C} - Z_{BC,\phi^A}\right)g^{\mu\nu}\phi^B_{,\mu}\phi^C_{,\nu} - 2Z_{AB}\lc{\square}\phi^B = 0\,.
\end{equation}
We conclude with the remark that although the form of the field equations is mostly unchanged compared to the single field case, there is an implicit and less apparent change coming from the fact that the parameter functions \(f\) and \(Z_{AB}\), and hence their derivatives appearing in the field equations, depend on all scalar fields.

\section{Summary and discussion}\label{sec:discussion}
In this paper we have presented a new class of theories in the covariant teleparallel framework, where the gravitational action depends on an arbitrary function of the torsion scalar and a scalar field, $f(T,\phi)$. This generic setup subsumes and generalizes a number of previously considered models, like $f(T)$ gravity and a scalar field nonminimally coupled to $T$, putting them in unified scheme so that they can be studied together. We derived the field equations for the tetrads, scalar field, and flat spin connection. The latter is especially important and until recently was missing in the covariant teleparallel picture. The spin connection equation turns out to be related to the antisymmetric part of the tetrad field equations and makes it to vanish identically. One also needs the spin connection equation when combining the field equations in order to show the matter energy-momentum conservation.

As a matter of fact, the spin connection field equation contains only first order derivatives with respect to the spacetime coordinates, and provides a consistency condition that from the tetrad ansatz determines the six nontrivial spin connection components (remaining after imposing zero curvature). These six components can be interpreted as gauge degrees of freedom, since they can be absorbed into the tetrad by a suitable local Lorentz transformation.

Solving the spin connection equations is not an easy matter, though. In a simple case the solution ansatz can reduce the equations to be those of the teleparallel equivalent to general relativity, where the spin connection can be fixed arbitrarily, but then the possibilities of the wider generalized theory remain unexplored. As we explain, for certain symmetric configurations it is possible to solve the spin connection equation independent of the function $f$, and illustrate this by the examples of cosmological and spherically symmetric spacetimes. These results can be used as a starting point for integrating the tetrad field equations, whereby one typically needs to specify the form of the function $f$. In the light of this understanding, not all previous results in $f(T)$ or scalar-torsion gravity can be automatically be taken with trust, one must check whether the assumed spin connection is consistent with the tetrad.

Our work leaves a number of possibilities for further investigations and generalizations. In particular it invites for studies of the phenomenology of the class of theories we discussed here, such as their post-Newtonian limit or gravitational waves. Also one may derive the cosmological field equations and perform an analysis of the possible solutions, employing the method of dynamical systems. Also fundamental questions, such as the number of propagating degrees of freedom, may be addressed, e.g., by performing a Hamiltonian analysis.

Another straightforward possibility is to consider more general action functionals, such as a Lagrangian given by an arbitrary function depending on the torsion scalar, the scalar field, its kinetic term and also involving a coupling to vector torsion~\cite{Hohmann:2018dqh}. Particular subclasses of such a model, where the scalar field couples to different terms constructed from the underlying teleparallel geometry by a small number of free functions, similar to the case of scalar-tensor gravity~\cite{Flanagan:2004bz}, are also worth studying~\cite{Hohmann:2018ijr}. One may also pose the question what is the most general theory coupling one or more scalar fields to torsion, and investigate its generic properties~\cite{Hohmann:2018vle}.

\begin{acknowledgments}
The authors thank Martin Kr\v{s}\v{s}\'ak and Christian Pfeifer for many useful discussions, as well as Ott Vilson, Tomi Koivisto and Alexey Golovnev for comments. The work was supported by the Estonian Ministry for Education and Science through the Institutional Research Support Project IUT02-27 and Startup Research Grant PUT790, as well as the European Regional Development Fund through the Center of Excellence TK133 ``The Dark Side of the Universe''.
\end{acknowledgments}

\bibliography{scaltors}
\end{document}